\documentclass[aps,prl,preprint,superscriptaddress,endfloats*,showpacs]{revtex4}
\usepackage{graphicx}
\usepackage{dcolumn}
\usepackage{bm}

\begin{document}

\title{ \hfill Phys.\ Rev.\ Lett. 97 (2006) \\
Real-Time {\em Ab Initio} Simulations of
        Excited Carrier Dynamics in Carbon Nanotubes}

\author{Yoshiyuki Miyamoto}
\affiliation{Fundamental and Environmental Research Laboratories,
             NEC Corp., 34 Miyukigaoka, Tsukuba,
             305-8501, Japan}

\author{Angel Rubio}
\affiliation
            {Dpto. F\'{\i}sica de Materiales, UPV/EHU,
             Centro Mixto CSIC-UPV/EHU, DIPC, 20018 San Sebasti\'an,
             Spain, and European Theoretical Spectroscopy Facility (ETSF)}

\author{David Tom\'anek}
\affiliation{Department of Physics and Astronomy,
             Michigan State University,
             East Lansing, Michigan 48824-2320}
\date{\today }

\begin{abstract}
Combining time-dependent density functional calculations for
electrons with molecular dynamics simulations for ions, we
investigate the dynamics of excited carriers in a $(3,3)$ carbon
nanotube at different temperatures. Following an $h\nu=6.8$~eV
photoexcitation, the carrier decay is initially dominated by
efficient electron-electron scattering. At room temperature, the
excitation gap is reduced to nearly half its initial value after
${\sim}230$~fs, where coupling to phonons starts dominating the
decay. We show that the onset point and damping rate in the phonon
regime change with initial ion velocities, a manifestation of
temperature dependent electron-phonon coupling.
\end{abstract}




\maketitle



Understanding the microscopic decay mechanism of electronic
excitations is a challenging problem in nanotechnology. The
fundamental hurdle is the requirement of unprecedented
computational resources to treat the time evolution of electronic
and ionic degrees of freedom in real time, from first principles,
for physically relevant time periods. Carbon nanotubes are the
ideal system to study these effects due to their well-defined
structure, intriguing electronic properties~\cite{book}, and a
high potential for application in future electronic
devices~\cite{HeinzePRL02}. Additional interest has recently been
triggered by observing optical emission induced by an electric
current~\cite{Misewich03}, radio-wave emission from aligned
nanotubes working as an antenna~\cite{WangAPL04}, and potential
application of nanotubes in a mode-locked laser~\cite{Tokumoto03}.
Whether studying the nature of excitonic effects affecting
photoabsorption~\cite{THeinz05}, or attempting to identify the
maximum switching frequency of future field effect transistors or
lasers, the key to progress is to understand the lifetime of
excited carriers in carbon nanotubes.

Time-resolved femtosecond pump-probe spectroscopy, applied to
nanotube samples, has identified two regimes in the decay of
carrier excitations~\cite{Hertel00}. An ultra-fast decay channel
has been associated with electron-electron scattering, which
causes an internal thermalization of the electronic system within
$200$~fs. A slower decay channel has been associated with the
excitation of ionic degrees of freedom at later times. Extracting
information about carrier decay in specific, isolated nanotubes
from those pump-probe data is not straight-forward, since samples
contain a mixture of metallic and semiconducting nanotubes with
different diameters and chiralities, all of which interact with
the substrate.

Here we present the first {\em ab initio} real-time calculation of
the decay mechanism of electronic excitations in narrow carbon
nanotubes, which exhibit strong electron-phonon
coupling~\cite{Benedict95,BohnenPRL04}. In our approach, we treat
the time dependence of the electronic degrees of freedom using
time-dependent density functional theory
(TDDFT)~\cite{Runge-Gross84}. Unlike in previous parameterized
TDDFT calculations \cite{Frauenheim05}, we determine concurrently
the ionic motion in the evolving charge density distribution by
direct integration using molecular dynamics (MD) simulations
within the force field given by the density functional theory
(DFT). Our results provide not only an unbiased interpretation of
available experimental data, but also identify the time scales
associated with the dominant carrier decay channels and their
temperature dependence. Our conclusion that the excitation gap is
reduced by electron-electron scattering first, followed by
electron-phonon interaction, is general and not restricted to
narrow carbon nanotubes.

Alternatively, selected aspects of excited carrier dynamics could
be described by summing up all diagrams corresponding to the
electron-electron and electron-phonon many-body interactions.
Under steady-state conditions, where the Fermi golden rule and
Boltzmann transport picture apply~\cite{PerebeinosPRL05},
electron-phonon coupling can be determined by integrating the
scattering matrix elements for transitions involving particular
phonon modes. The corresponding scattering time constant,
including its temperature dependence, could then be determined
using the inverse momentum-space integrals of these matrix
elements~\cite{Jiang05}. This steady-state approach, however, is
not applicable in the sub-picosecond regime of interest here,
since electron-electron scattering processes have been averaged
out. Dynamics of electronic decay channels only, ignoring coupling
to phonons, could also be inferred from the imaginary part of the
electron self-energy, determined using the GW approximation
\cite{Spataru01}.

Unlike these approaches, our calculations offer an unbiased
insight into the short-time dynamics of electronic excitations,
since electron and phonon scattering are treated on the same
footing in real time within TDDFT-MD, using the FPSEID (First
Principles Simulation tool for Electron Ion Dynamics)
code~\cite{Sugino-Miyamoto99}. Since the systems of interest are
well above zero temperature, the ionic motion is described
classically, with the forces acting on the ions given by the
density functional theory in the local density approximation
(LDA)~\cite{PZ}. The electron-ion interaction is described using
norm-conserving soft pseudopotentials~\cite{TM}, and the valence
wave functions are expanded in a plane wave basis with a kinetic
energy cutoff of $40$~Ry.

\begin{figure}[t]
\includegraphics[width=0.83\columnwidth]{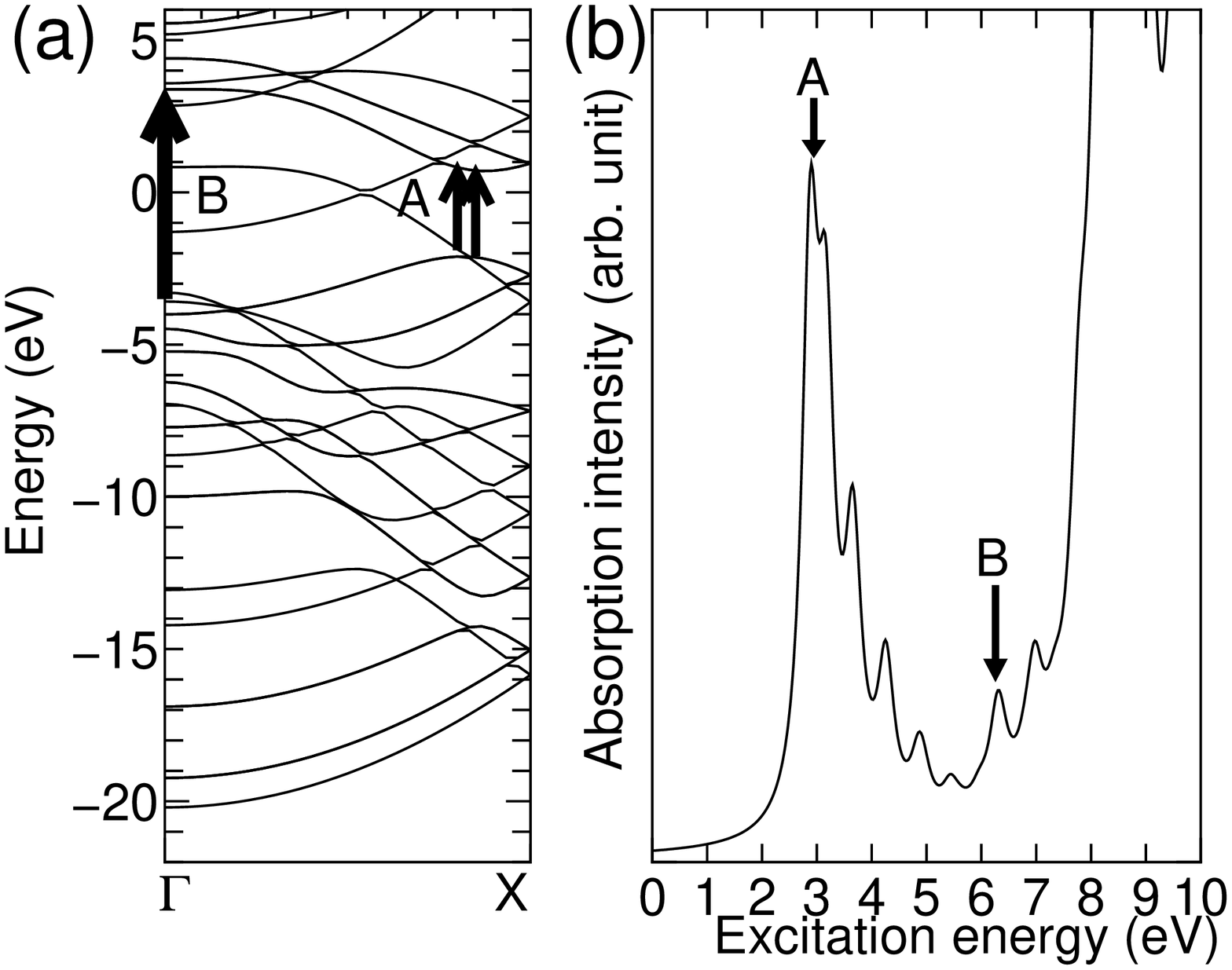}
\caption{ (a) LDA band structure and (b) optical absorption
spectrum of an isolated $(3,3)$ nanotube for a dipole field along
the tube axis. Two dipole-allowed transitions are labelled ``A''
and ``B''. The Fermi level lies at $0$~eV. The absorption spectrum
(b) has been convoluted with a $0.3$~eV wide Lorentzian.}
\label{Fig1}
\end{figure}

To allow for a realistic representation of phonons in isolated
$(3,3)$ nanotubes, we arrange 96-atom supercells, which extend
over eight primitive unit cells of the tube, in a hexagonal
geometry with an inter-wall separation of 5~{\AA}. We find that
$\Gamma$ point sampling of the small Brillouin zone is sufficient
to describe the electron-electron and electron-phonon coupling
well. To describe the dynamics at a given temperature, we
randomize the initial ion velocities according to the
Maxwell-Boltzmann distribution for that thermodynamical ensemble.
The excited-state dynamics then proceeds in three steps. First, we
simulate the initial optical excitation by promoting a valence
electron to an unoccupied Kohn-Sham (KS) state. Next, by fixing
this electronic configuration, we use constrained DFT to determine
the electronic structure in that excited state. Finally, we
continue a full, unconstrained TDDFT-MD simulation starting with
these initial ion velocities and charge distribution.

Figure~\ref{Fig1} depicts the electronic structure and
photoabsorption spectrum in the narrow $(3,3)$ carbon
nanotube~\cite{narrowtubes-expt}. The electronic band structure of
the primitive unit cell, shown in Fig.~\ref{Fig1}(a), agrees with
published data~\cite{Li01,MarinopoulosPRL03} also in displaying
the curvature effect, which modifies the Fermi momentum from the
zone folding value at $2/3({\Gamma}-X)$. The photoabsorption
spectrum in Fig.~\ref{Fig1}(b) displays the response to light with
the E-field polarized along the tube axis. Since depolarization
effects normal to the tube are not important due to the small
cross-section~\cite{MarinopoulosPRL03}, we used Fermi's golden
rule directly to calculate the optical dipole matrix elements
between occupied and empty KS
eigenstates~\cite{spectrum-corrections}.

We found two relevant photoabsorption processes, labelled ``A''
and ``B'' in Fig.~\ref{Fig1}. The absorption spectrum is dominated
by a peak labeled ``A''~\cite{Li01}, which stems from many
independent electron-hole transitions in a narrow energy range. A
complex competition between the electronic decay channels, each
with a different time constant, and phonon decay channels makes a
clear distinction between electronic and phonon contributions
problematic and will not be attempted here. In the following, we
rather focus on the transition labelled ``B'' in Fig.~\ref{Fig1},
which, in spite of its smaller oscillator strength, is better
defined due to the absence of nearby transitions with a large
oscillator strength and allows a clear distinction between
electronic and ionic decay channels. Our findings in terms of
dominant decay channels, we believe, are of general nature,
applicable to other photoabsorption processes in related systems.

\begin{figure}[t]
\includegraphics[width=0.70\columnwidth]{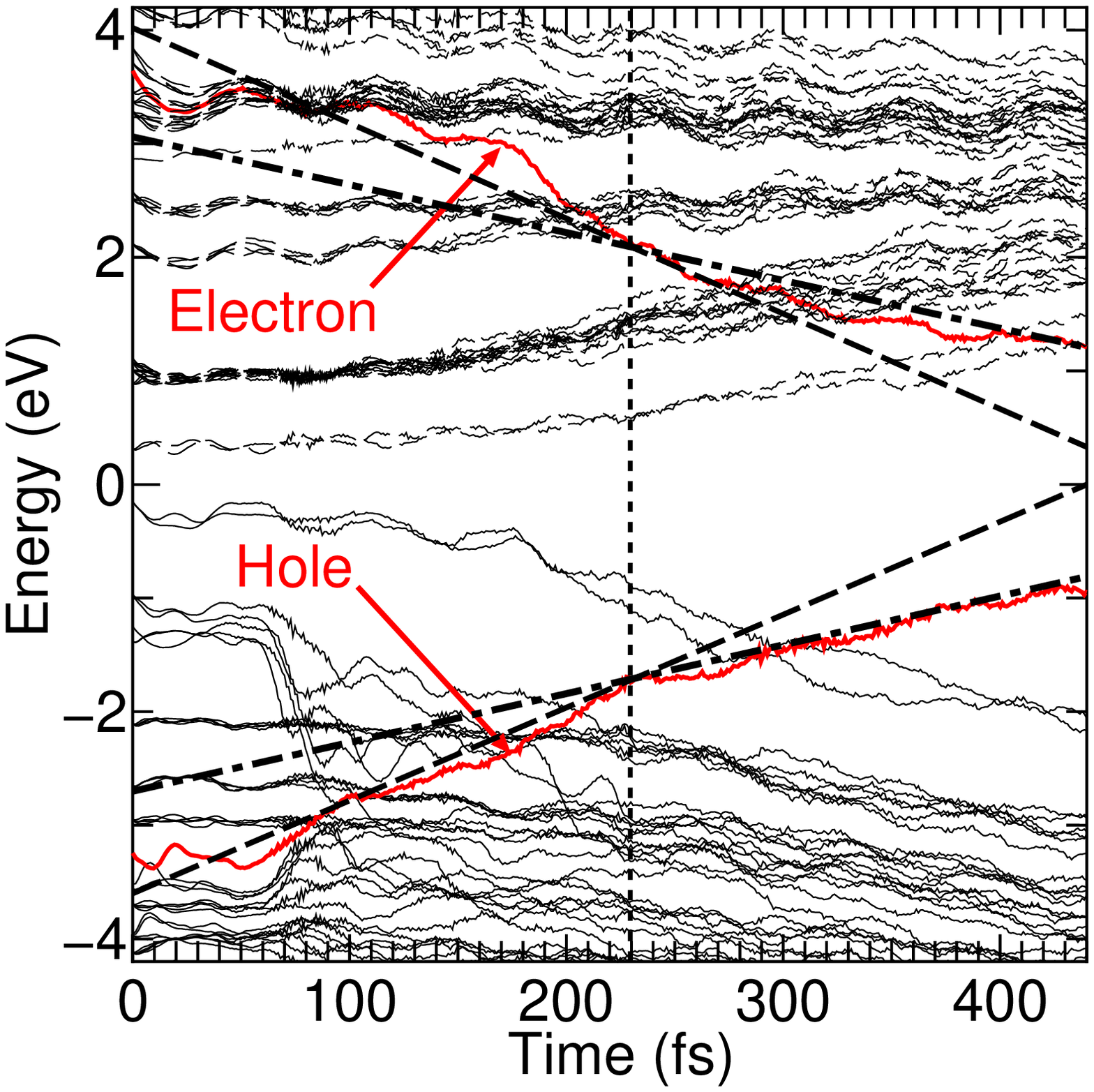}
\caption{ (Color online) Dynamics of hot-carrier decay following
the excitation labelled ``B'' in Fig.~\ref{Fig1}, after the ions
have been thermalized at $T=300$~K. Valence bands are shown by
solid lines and conduction bands by dotted lines. The excited
electron and hole states are shown by heavy solid lines. Dashed
lines are guides to the eye. } \label{Fig2}
\end{figure}

The time evolution of the electronic spectrum following a ``B''
excitation at $t=0$ is shown in Fig.~\ref{Fig2}. The time
evolution of the KS eigenvalues indicates a rapid reduction of the
electron-hole excitation gap~\cite{excitation-gap} to less than
half its initial value of $6.8$~eV within $500$~fs from the
photoexcitation. The individual electron (hole) eigenvalues show
an initial tendency to reach the bottom (top) of the conduction
(valence) band very rapidly, manifesting the efficient
non-radiative decay of the optical excitation towards the lowest
lying electronic excitation. Under room temperature conditions,
depicted in Fig.~\ref{Fig2} and achieved by initially thermalizing
the ionic motion at $T=300$~K, we find a noticeable reduction of
the decay rate ${\sim}230$~fs after the photoexcitation.
Subsequent decay may involve less efficient radiative
transitions~\cite{THeinz05} or energy transfer to phonons.

We believe that the predicted reduction of the electron-hole
energy gap could in principle be observed in time-resolved
experiments. Light emission as a result of electron-hole
recombination could generally be expected, since the computed
optical matrix elements between the excited electron and hole
states fluctuate for a long time around a finite value. Even
though non-radiative decay of the excitation towards the lowest
excitonic state is typically faster than radiative decay, emission
could possibly be resonantly enhanced by the probe laser.

To gain additional insight into the electronic decay, including
possible radiative transitions, we investigated the time evolution
of the matrix containing electron-hole dipole matrix elements.
Immediately after the photoexcitation, the only non-vanishing
matrix elements involve a field along the tube axis, same as in
the initial photo-absorption. Later on, as phonons weakened the
dipole selection rules, we observed matrix elements involving a
field normal to the tube axis to increase, and eventually to reach
similar values as those involving a field along the tube axis. As
the electron-hole dipole matrix elements become isotropic in time,
also the radiative decay should not show any preferential
polarization.

In spite of substantial changes in its electronic structure, the
$(3,3)$ nanotube retains its cylindrical shape throughout the
simulation. Close inspection of the vibrational spectra indicates
enhancement of particular phonon modes following the
photoexcitation, in particular at $\omega=556$~cm$^{-1}$, close to
the $\omega_{RBM}=536$~cm$^{-1}$ value reported for the the radial
breathing mode (RBM) of the $(3,3)$ carbon
nanotube~\cite{BohnenPRL04}. Our conclusion that the RBM may be
excited by photoabsorption is also supported by recent femtosecond
pump-probe experiments, which report exciting coherent phonons,
including the RBM, in semiconducting single wall carbon nanotubes
using resonant sub-10~fs pulses in the visible
range~\cite{Gambetta05}.

\begin{figure}[t]
\includegraphics[width=1.0\columnwidth]{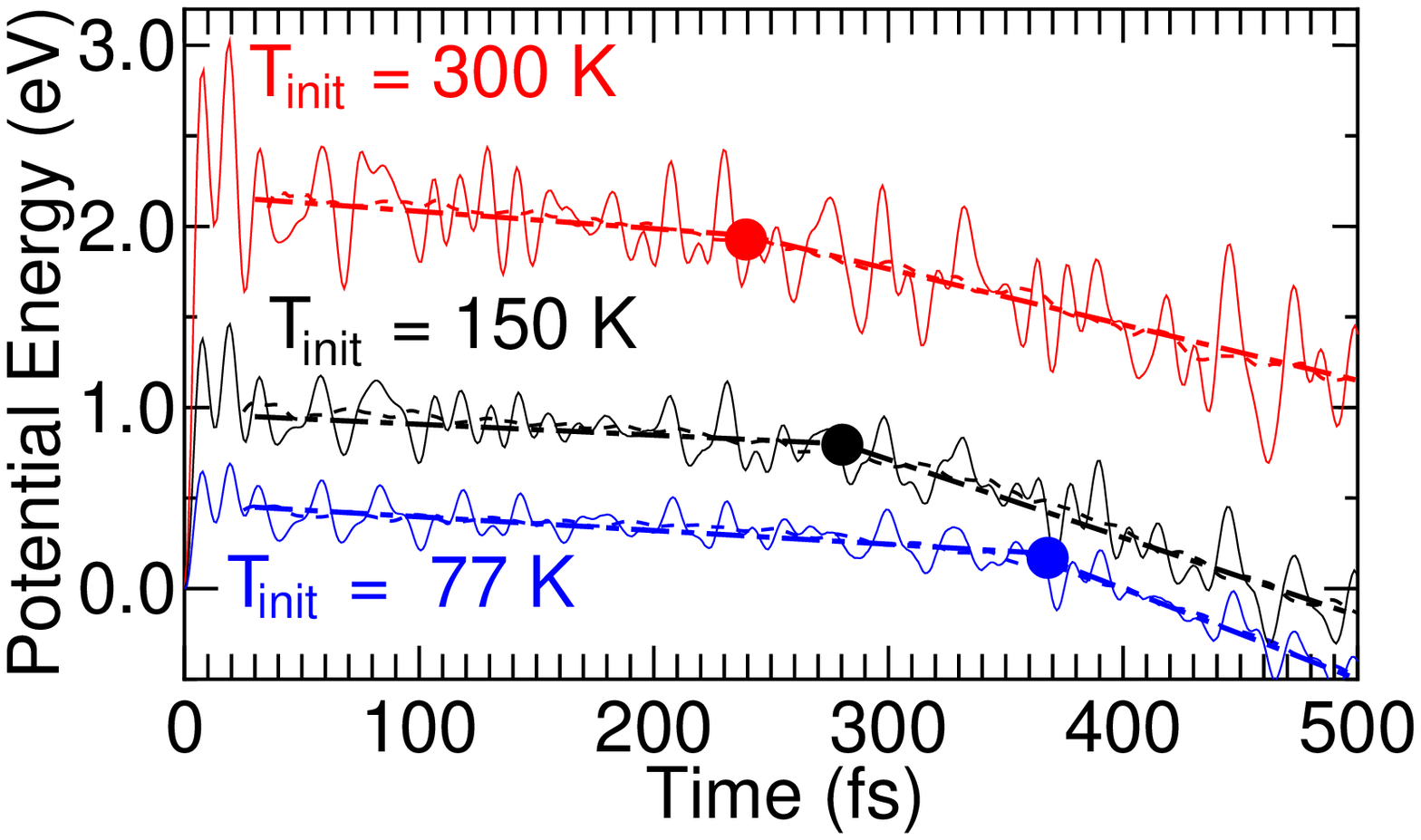}
\caption{ (Color online) Time dependence of the ionic potential
energy after the photoexcitation at $t=0$. The different solid
lines describe systems with an initial Maxwell-Boltzmann velocity
distribution corresponding to $T=77$~K, $150$~K, and $300$~K. The
light dotted lines represent time-averaged values. The dashed
lines, showing linear fits to the data, are guides to the eye to
distinguish the two time regimes. \label{Fig3}}
\end{figure}

The time-dependence of the photoexcitation gap, discussed in
conjunction with Fig.~\ref{Fig2}, tells little about the nature
and temperature dependence of the processes controlling the hot
carrier decay. To obtain this information, we investigated the
photo-decay in nanotubes with initial ionic temperatures of
$T=77$~K, $150$~K, and $300$~K. Witnessing to the numerical
long-time stability of our TDDFT-MD simulations, we find the total
energy to be conserved with a maximum deviation of about
$10^{-3}$~eV/atom over the $0.5$~ps long simulations.
Figure~\ref{Fig3} depicts the key result, namely the time
dependence of the potential energy of the nanotube under different
initial conditions.

Due to the finite size of our unit cell, the potential energy
fluctuates in time, with larger fluctuations occurring at higher
temperatures. Since ensemble averaging over may runs would
necessitate unreasonably large computer resources, we invoked
ergodicity and time averaged our potential energy profiles over
$50$~fs long time periods. Our results, displayed by the dotted
lines in Fig.~\ref{Fig3}, indicate that independent of the initial
temperature, the potential energy decreases in time. After
reaching a cross-over point, an initially gradual decrease becomes
steeper. We find the decay rate to be nearly linear in both
regimes, as shown by the dashed lines in Fig.~\ref{Fig3}. Whereas
the decay rate in the early regime is rather independent of the
initial temperature, its length decreases from ${\sim}360$~fs at
$77$~K to ${\sim}230$~fs at $300$~K. Also, in the regime beyond
the cross-over point, the decay rate decreases with increasing
temperature, suggesting a slower energy flow from the electronic
system to the ionic lattice. This is a clear indication of a
temperature-dependent electron-phonon coupling.

In contrast to the time dependence of the potential energy
discussed above, the electron-hole energy gap, shown in
Fig.~\ref{Fig2}, first shows a rapid, then a more gradual
decrease. The cross-over points between the two decay regimes at
$300$~K, based on the electronic spectra in Fig.~\ref{Fig2} and
the potential energy in Fig.~\ref{Fig3}, nearly coincide at
$t{\sim}230$~fs, suggesting the same underlying mechanism. Our
finding that the initial decay is fast in Fig.~\ref{Fig2}, yet
slow and temperature independent in Fig.~\ref{Fig3}, suggests that
early decay regime is dominated by relaxations within the
electronic system only. The opposite behavior is observed in the
second regime, suggesting involvement of phonons. In view of the
stronger electron-phonon coupling in the narrow $(3,3)$ nanotube,
our estimated cross-over point between the regimes should still
remain valid as a lower limit for wider nanotubes. In view of
uncertainties associated with samples containing mixtures of
nanotubes, the value $t{\sim}200$~fs, observed~\cite{Hertel00} in
single-wall nanotubes with diameters close to $1.2$~nm, lies very
close to our prediction.

Even though the time evolution of the charge density, underlying
the density functional force field on the ions, is based on
Ehrenfest dynamics~\cite{Ehre27}, our calculated forces do reflect
the electron and ion dynamics at particular temperatures without
relying on the adiabatic potential energy surface for a particular
excited state. Deviations from Ehrenfest dynamics only occur, when
the system cannot be associated with the adiabatic potential
energy surface of an individual excited state. This would give
rise to non-vanishing off-diagonal matrix elements of the
time-dependent Kohn-Sham matrix, evaluated in the basis of the
corresponding time-dependent eigenstates. We have indeed observed
such non-vanishing off-diagonal matrix elements, albeit very small
in magnitude, and only during short time periods in our
simulation~\cite{epaps}. As another point worth mentioning, the electron-hole
attraction in excitonic states, which is not reproduced by the
DFT, is known not to change the total charge density much and thus
not to affect the electron and ion dynamics. Thus, in stark
contrast to the strong influence of ions on the excited carriers,
which we consider explicitly, we find the influence of electronic
excitations on the ion dynamics to be negligible. All the above
arguments, along with the high degree of total energy conservation
in our system discussed earlier, justify our use of Ehrenfest
dynamics.


In conclusion, our large-scale {\em ab initio} simulations of
photophysical processes in carbon nanotubes suggest that the decay
of photoexcited carriers occurs on a sub-picosecond time scale. An
initial rapid nonradiative electronic decay is followed by a
temperature-dependent, slower decay involving phonons. The
duration of the electronic decay regime and the decay rate in the
phonon-dominated regime decrease with increasing temperature,
suggesting temperature-dependent electron-phonon coupling.


YM acknowledges fruitful discussions with Dr. B.D. Yu and major
use of the Earth Simulator Supercomputer. AR was supported by the
EC projects NANOQUANTA (NMP4-CT-2004-500198), SANES
(NMP4-CT-2006-017310), Spanish MCyT, and the Humboldt Foundation.
DT was supported by NSF NIRT grants DMR-0103587, ECS-0506309, NSF
NSEC grant EEC-425826, and the Humboldt Foundation.



\begin{thebibliography}{}
\bibitem{narrowtubes-expt} Even though free-standing $(3,3)$
         nanotubes have not been reported, calculations indicate
         that they are stable. Optical spectroscopy also confirms
         that such narrow nanotubes may form inside pores of a
         zeolite~\protect\cite{Li01}.
\bibitem{spectrum-corrections} The photoabsorption spectrum was
         obtained by sampling the Brillouin zone
         with 16 k-points. For quantitative agreement with
         experiment, the spectrum in Fig.~\protect\ref{Fig1}(b)
         should include self-energy and excitonic effects
         which are beyond the scope of this work.
\bibitem{excitation-gap} Taking total energy differenes based on
         constrained LDA calculations, we found the electron-hole
         excitation gap corresponding to transition ``B'' to lie
         at $6.89$~eV, close to the Koopman's Theorem value of
         $6.8$~eV.
\end{thebibliography}
\end{document}